\documentclass[ reprint,
 superscriptaddress,
 amsmath,amssymb,
 aps,
 pra,
 showkeys,
 nofootinbib
]{revtex4-2}

\usepackage{overpic}
\usepackage{booktabs}
\usepackage{framed}
\clearpage{}\usepackage[uline]{hhtensor}
\usepackage{mathrsfs}
\usepackage{hyperref}
\usepackage{mathtools}
\usepackage{siunitx}
\usepackage{xcolor}
\usepackage{textgreek}
\newcommand*{\hbaraux}[2]{\sbox0{\mathsurround=0pt$#1\mkern-1mu\mathchar'26$}\mkern-1mu\lower.07\ht0\box0\mkern-8mu}
\renewcommand*{\hbar}{{\mathpalette\hbaraux\relax\mathrm{h}}}

\newcommand\duetoeq[1]{\stackrel{\text{Eq.\ (\ref{#1})}}{=}}

\newcommand\equaldueto[1]{\stackrel{#1}{=}}
\newcommand\equalduetoeq[1]{\stackrel{\text{Eq.~(\ref{#1})}}{=}}

\newcommand\Eq[1]{\text{Eq.~(\ref{#1})}}

\newcommand\Phistaticket{|\Phi_{\omega=0}\rangle}

\newcommand\rr{\mathbf{r}}

\newcommand\pp{\mathbf{k}}

\newcommand\phat{\mathbf{\hat{\pp}}}

\newcommand\PP{\mathbf{P}}
\newcommand\JJ{\mathbf{J}}

\newcommand\Help{\ii\phat\times}

\newcommand\Fpplus{\mathbf{F}_+(\pp)}
\newcommand\Fpminus{\mathbf{F}_-(\pp)}
\newcommand\Gpplus{\mathbf{G}_+(\pp)}
\newcommand\Gpminus{\mathbf{G}_-(\pp)}

\newcommand\sphharm[3]{\mathrm{Y}_{#1}^{#2}(#3)}

\newcommand\ff[1]{\mathrm{f}_{#1}(\pp)}
\newcommand\FF[1]{\mathrm{f}_{jm#1}(|\pp|)}

\newcommand\Frtplusminus{\mathbf{F}_{\pm}(\rr,t)}

\newcommand\rhort{\rho(\rr,t)}

\newcommand\Ert{\mathbf{E}(\rr,t)}

\newcommand\Brt{\mathbf{B}(\rr,t)}

\newcommand\Id{\mathrm{I}}

\newcommand\intdpconfhbar{\int_{\mathbb{R}^3} \frac{\mathrm{d}^3 \pp}{\hbar\cz|\pp|}\text{ }}

\newcommand\intdpM{\int_{\mathbb{R}^3}\frac{\muz \mathrm{d}^3\pp}{\cz\hbar|\pp|}}
\newcommand\intdpMsp{\int_{\mathbb{R}^3}\frac{\mathrm{d}^3\pp}{|\pp|}}
\newcommand\intdpMnohbar{\int_{\mathbb{R}^3}\frac{\muz \mathrm{d}^3\pp}{\cz|\pp|}}
\newcommand\intdmodpM{\int_{0}^\infty \ \mathrm{d}|\pp||\pp|}

\newcommand\intdpnorm{\int_{\mathbb{R}^3} \frac{\mathrm{d}^3 \pp}{\sqrt{(2\pi)^3}}\text{ }}
\newcommand\intdpnormV{\int_{\mathrm{V}} \frac{\mathrm{d}^3 \rr}{\sqrt{(2\pi)^3}}\text{ }}

\newcommand\intdr{\int_{\mathbb{R}^3} {\mathrm{d}^3 \rr}\text{ }}

\newcommand\zerovec{\mathbf{0}}
\newcommand\mrest{\mathbf{M}(\rr)}
\newcommand\Mr{\mathbf{M}(\rr)}

\newcommand\mhatr{\mathbf{\hat{m}}(\rr)}

\newcommand\rhorest{\rho(\rr)}

\newcommand\ii{\mathrm{i}}

\newcommand\Jrt{\mathbf{J}(\rr,t)}

\newcommand\Mrt{\mathbf{M}(\rr,t)}
\newcommand\Prt{\mathbf{P}(\rr,t)}

\newcommand\muz{\text{\textmu}_{\text{0}}}
\newcommand\epsz{\text{\textepsilon}_{\text{0}}}
\newcommand\cz{\mathrm{c_0}}

\newcommand\Er{\mathbf{E}(\rr)}

\newcommand\Br{\mathbf{B}(\rr)}

\newcommand\Ar{\mathbf{A}(\rr)}

\newcommand\Hr{\mathbf{H}(\rr)}

\newcommand\Mp{\mathbf{M}(\pp)}

\newcommand\Fplambda{\mathbf{F}_\lambda(\pp)}

\DeclareMathOperator{\atantwo}{atan2}

\newcommand{\mhatthetazero}{\mathbf{\hat{m}}_{\theta=0}}

\newcommand{\Qlambda}{\mathbf{Q}_\lambda(\phat)}

\newcommand\rhor{\rho(\rr)}
\newcommand\RR{\mathrm{R}}

\clearpage{}
\usepackage{empheq}

\renewcommand\Eq[1]{\text{Equation~(\ref{#1})}}
\renewcommand\duetoeq[1]{\stackrel{\text{Equation\ (\ref{#1})}}{=}}
\renewcommand\equalduetoeq[1]{\stackrel{\text{Equation~(\ref{#1})}}{=}}

\begin{document}
\title{A scalar product for computing fundamental quantities in matter}
\author{Ivan Fernandez-Corbaton}
\email{ivan.fernandez-corbaton@kit.edu}
\affiliation{Institute of Nanotechnology, Karlsruhe Institute of Technology, 76021 Karlsruhe, Germany}
\author{Maxim Vavilin}
\affiliation{Institute of Theoretical Solid State Physics, Karlsruhe Institute of Technology, 76131 Karlsruhe, Germany}
\begin{abstract}
	We introduce a systematic way to obtain expressions for computing the amount of fundamental quantities such as helicity and angular momentum contained in static matter, given its charge and magnetization densities. The method is based on a scalar product that we put forward, which is invariant under the ten-parameter conformal group in three-dimensional Euclidean space. Such group is obtained as the static restriction (frequency $\omega=0$) of the symmetry group of Maxwell equations: The fifteen-parameter conformal group in 3+1 Minkowski spacetime. In an exemplary application, we compute the helicity and angular momentum squared stored in a magnetic Hopfion.
\end{abstract}


\maketitle

\section{Introduction and summary}
Research in light--matter interactions benefits from theoretical tools whose generality allows one to treat light and matter in similar ways. Symmetries and conservation laws are prominent examples of such generic tools \cite{Wigner1959}. For example, linear and angular momentum are physical quantities that are tied to symmetry transformations and that apply to both light and matter. These quantities have in common that they are the generators of transformations in symmetry groups that are relevant in physics, such as the Poincar\'e group of special relativity \cite{Wigner1939}.

In classical electrodynamics, light and matter are treated in a rather similar way, namely, as continuous fields, such as the electric and magnetic fields $\Ert$ and $\Brt$ representing radiation, and the densities of charge $\rhort$, current $\Jrt$, polarization $\Prt$, and magnetization $\Mrt$, representing matter. Light can be treated by using the tools of Hilbert spaces \cite{Gross1964,Moses1973b,Birula1975,Birula1996,FerCor2016c}, facilitating the consideration of material symmetries and their consequences for light upon light--matter interaction \cite{FerCor2016c}. The conformal invariance of Maxwell equations \cite{Bateman1910,Cunningham1910,Dirac1936} and the corresponding conformally invariant scalar product between free electromagnetic fields \cite{Gross1964,Birula1975,Birula1996} play an important role in such algebraic approach to electrodynamics. In particular, the scalar product allows one to obtain expressions for computing the amount of quantities such as angular momentum, energy, and helicity contained in a given radiation field \cite{Birula1996,Birula1975,FerCor2019b}. For the radiation field, helicity is essentially its polarization handedness.

In this article, we extend the algebraic approach towards the matter side and introduce a way to obtain expressions for computing the amount of fundamental quantities contained in static material objects from their charge and magnetization densities, $\rhor$ and $\Mr$, respectively. The densities are assumed to be spatially confined and real--valued. The expressions are obtained as a scalar product $\langle\Phi|\Gamma|\Phi\rangle$, where $|\Phi\rangle$ represents $\left\{\rhor,\Mr\right\}$, and $\Gamma$ is the self--adjoint (Hermitian) operator representing the particular quantity of interest. The expression of the scalar product for static matter is derived for the static fields that are bijectively connected to $\rhor$ and $\Mr$. To this end, the largest group of transformations that leave Maxwell equations with sources invariant, that is, the fifteen-parameter conformal group in 3+1 Minkowski spacetime \cite{Bateman1910,Cunningham1910} $C_{15}(3,1)$ is considered first, then the transformations that would not preserve the frequency $\omega=0$ condition of static fields are excluded. The remaining transformations also form a group, the ten-parameter conformal group in three-dimensional Euclidean space $C_{10}(3)$ \cite[Chapter~24]{Grafarend2014}. We put forward an expression for a scalar product and show that it is invariant under $C_{10}(3)$. In particular, it allows for explicit numerical computations of $\langle\Phi|\Gamma|\Phi\rangle$. We discuss the importance of invariant scalar products for the consistent definition of measurements and quantities such as $\langle\Phi|\Gamma|\Phi\rangle$, which provides the motivation for finding a relevant group of transformations and an invariant scalar product for the static case.

In an exemplary application, we set $\rhor=0$ to focus on magnetization textures and compute the helicity and total angular momentum squared stored in a Hopfion \cite{Sutcliffe2018,Liu2018,Jung2018,Kent2021,Khodzhaev2022} hosted inside a FeGe disk under zero external field. The definition of helicity that we use in this article [\Eq{eq:Hel}] is the one commonly used in optics and field theory, and can be understood as the sense of screw. This is {\em not the same} as the definition often used for magnetic solitons \cite{Verba2020}. One important difference is that the latter can be defined for two-dimensional objects, while the former needs three spatial dimensions. With the definition in \Eq{eq:Hel}, the helicity of the Hopfion is equal to -129.1$\hbar$. This number is a lower bound for the number of circularly polarized photons of positive helicity that would be needed in a helicity--dependent all--optical switch of the Hopfion to its mirror image of opposite handedness: \protect{$\lceil 129.1\times 2 \rceil=259$}. The number \protect{-129.1$\hbar$} also bounds the helicity that can be radiated from the Hopfion as it loses its chirality, for example by the action of a large magnetic bias aligning its magnetization density vector along the same direction at all points in the magnet. 

More generally, we find that the net angular momentum and linear momentum of any $\{\rhor,\Mr\}$ along any given axis vanishes, which is consistent with the assumption of static matter.

The rest of this article is organized as follows. Section~\ref{sec:mot} introduces the setting and context of the work. The relevance of invariant scalar products for the consistent definition of measurements and quantities such as $\langle\Phi|\Gamma|\Phi\rangle$ is discussed in Section~\ref{sec:measurement}. The scalar product for matter in static equilibrium is presented in Section~\ref{sec:sp}, where its conformal invariance under $C_{10}(3)$ is shown, providing the basis for a new approach to computing properties of static material objects of finite volume from their charge and magnetization densities. In Section~\ref{sec:form} we provide explicit formulas for computing the helicity, angular momentum, and angular momentum squared of a static magnetization texture $\Mr$ and show that the net angular momentum vanishes along any given axis. In Section~\ref{sec:hopfion}, we use the presented formalism for a quantitative study of a Hopfion in a FeGe disk. Section~\ref{sec:concl} concludes the article. 

The approach presented here makes the computation of fundamental quantities in matter very similar to corresponding computations for the electromagnetic field, and can be readily applied to analytically derived \cite{Sutcliffe2018}, numerically obtained \cite{Peralta2007}, or experimentally measured three-dimensional charge \cite{Coppens2005} and magnetization density distributions \cite{Donnelly2017}. We foresee that the methodology will particularly be useful for the design and analysis of experiments involving the switching between stable states of a material system, such as, for example, when using circularly polarized light to switch the magnetization direction in magnetic films \cite{Stanciu2007}, which indicates a path towards much faster and more energy efficient computer memories, and whose underlying mechanisms are under intense scrutiny.

\section{Motivation and problem setting\label{sec:mot}}

\begin{figure}[h]
		\begin{overpic}[width=\linewidth]{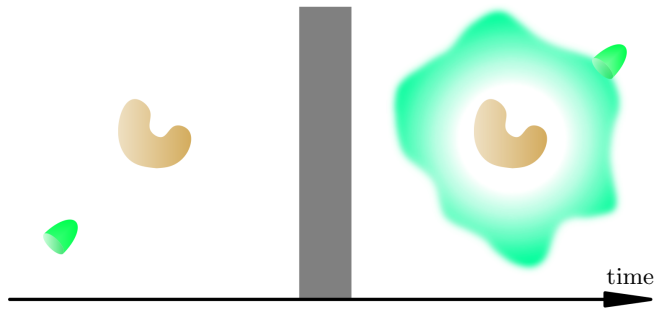}
							\put(92,5){time}
	\end{overpic}

	\caption{A beam of electromagnetic radiation interacts with a material object of finite size during the grayed--out period. Before and some time after the interaction the object is in static equilibrium, where the time derivatives of all macroscopic quantities vanish. The interaction typically changes fundamental quantities of the field such as its energy or momentum. Well known formulas exist for computing such quantities for the electromagnetic field. In this article, we develop a method to compute them for the material object in static equilibrium.\label{fig:lmi}}
\end{figure}

Figure~\ref{fig:lmi} depicts a light--matter interaction sequence. A beam of electromagnetic radiation approaches a material object of finite size. Before the start of the interaction the object is in equilibrium with the radiation field. Then, the beam and the object interact for a finite period of time. When the interaction stops and equilibrium is reached again, both the beam and the object may have changed. For example, the energy, momentum, and angular momentum contained in the radiation field before and after the interaction may be different. The same can be said about the material system. 

There are well--known expressions for computing the amount of a fundamental quantity such as energy or momentum contained in a given electromagnetic field. For example, in SI units which are used throughout this article, and with $\epsz$ and $\muz$ respectively denoting the permittivity and permeability of vacuum, we have 
\begin{equation}
	\label{eq:hpdr}
		\frac{1}{2}\intdr \left(\epsz\mathbf{\mathcal{E}}\cdot\mathbf{\mathcal{E}}+\frac{1}{\muz}\mathbf{\mathcal{B}}\cdot\mathbf{\mathcal{B}}\right)\text{ and } \epsz\intdr \mathbf{\mathcal{E}}\times\mathbf{\mathcal{B}}
\end{equation}
as the energy and momentum of the field, respectively. The fields $\mathbf{\mathcal{E}}$ and $\mathbf{\mathcal{B}}$ are real--valued. As is explained later, we will use complex fields with positive frequencies to describe the dynamic electromagnetic field, which we denote by $\mathbf{E}$ and $\mathbf{B}$, and for which the relation with the real--valued fields is $\mathbf{\mathcal{X}}=\mathbf{X}+\mathbf{X}^*$.

Expressions such as those in \Eq{eq:hpdr} can be derived in several different ways. For example, using conservation laws as in \cite[Chapter~3]{Schwinger1998}, or integrating the electromagnetic stress tensor as in \cite[Chapter~12.10]{Jackson1998}. An alternative approach uses the tools of Hilbert spaces. In such a framework, the fields are vectors in the Hilbert space of free solutions of Maxwell equations, that is, electromagnetic fields that are not interacting with matter. Each particular solution $\{\Ert,\Brt\}$ corresponds to a ket $|\Phi\rangle$. The fundamental quantities are represented by self--adjoint operators that act on the kets. Then, the total amount of a given fundamental quantity $\Gamma$ contained in a given electromagnetic field $|\Phi\rangle$ can be written as the scalar product of $\Gamma|\Phi\rangle$ and $|\Phi\rangle$:
\begin{equation}
	\label{eq:gamma}
	\langle\Phi|\Gamma|\Phi\rangle.
\end{equation}

The expression of the scalar product for the free radiation fields, which in particular is used for obtaining explicit expressions from \Eq{eq:gamma}, reads \cite{Gross1964,Birula1996}:
\begin{equation}
	\label{eq:scalarproduct}
	\langle \text{F}|\text{G}\rangle=\intdpconfhbar \begin{bmatrix}\Fpplus\\\Fpminus\end{bmatrix}^\dagger \begin{bmatrix}\Gpplus\\\Gpminus\end{bmatrix},
\end{equation}
	where $\cz=1/\sqrt{\epsz\muz}$, and the two kets $|\text{F}\rangle$ and $\text{G}\rangle$ are represented by their plane wave components of well--defined helicity $\lambda=1$ and $\lambda=-1$, $\Fpplus$ and $\Fpminus$, respectively, which correspond to left-- and right--handed circular polarizations:
\begin{equation}
	\label{eq:frt}
	\begin{split}
		&\Frtplusminus=\sqrt{\frac{\epsz}{2}}\left[\Ert\pm\ii \cz\Brt\right]\\&=\intdpnorm \Fplambda\exp(i\pp\cdot\rr-\ii\cz|\pp| t),
	\end{split}
\end{equation}
	with $\pp\cdot\Fplambda=0$, and, importantly, the angular frequency is restricted her to positive values $\omega=\cz|\pp|>0$. The exclusion of $\omega<0$ is possible in electromagnetism because both sides of the spectrum contain the same information \cite[\S 3.1]{Birula1996}\cite{Birula1981}. The $\omega=0$ point is also excluded from the domain of the dynamic fields.

	The $\Frtplusminus$[$\Fplambda$] are eigenstates of the helicity operator $\Lambda$. Helicity is the projection of the angular momentum $\JJ$ onto the direction of the linear momentum $\PP$. The $\pp$--space representation of $\Lambda$ is particularly simple:
	\begin{equation}
		\label{eq:Hel}
		\begin{split}
			\Lambda=\frac{\JJ\cdot\PP}{|\PP|}&\equiv \hbar\Help,\\
			\hbar\Help\mathbf{F}_\pm(\pp)&=\pm\hbar\mathbf{F}_\pm(\pp).
		\end{split}
	\end{equation}

	The defining property of the scalar product in \Eq{eq:scalarproduct} is that it is conformally invariant \cite{Gross1964}. That is, the value of $\langle \text{F}|\text{G}\rangle$ is identical to the scalar product between $\text{X}|\text{F}\rangle$ and $\text{X}|\text{G}\rangle$, for any transformation $\text{X}$ in the conformal group in 3+1 Minkowski spacetime. This group is the largest group of invariance of Maxwell equations {\em including sources as spacetime densities} \cite{Bateman1910,Cunningham1910}. In particular, free Maxwell fields transform into free Maxwell fields under the conformal group \cite{Gross1964}. The group consists of spacetime scalings, four special conformal transformations, and the Poincar\'e group, which consists of four spacetime translations, three Lorentz boosts, and three spatial rotations \cite{Budinich1993,Fuschchich1994}. 

	The quantity under the integral sign in \Eq{eq:scalarproduct} is unitless. This can be verified by the direct computation of its units, where it should be take into account that, according to \Eq{eq:frt}, the units of $\Fplambda$ are equal to the units of $\Frtplusminus$ times $\SI{}{\meter}^3$. A conformally invariant scalar product must be unitless, because spacetime scalings and special conformal transformations can be interpreted as changes of units \cite{Kastrup1962,Barut1972,Kastrup2008}. In the case of the spacetime scalings, the change of units is the same for all spacetime points, while in the case of the special conformal transformations, the change varies with space and time.

	Together, \Eq{eq:gamma} and \Eq{eq:scalarproduct} are a general and convenient way of computing the amount of any fundamental quantity in the field. This motivates the following question: {\em Can this algebraic approach be used for the material object?}

An affirmative answer to this question requires an appropriate mathematical representation of matter, an appropriate group of transformations, and a scalar product that is invariant under all the group transformations. These latter two requirements are explained in the following section.

	\section{The importance of invariant scalar products in the consistent definition of measurements \label{sec:measurement}}
	The conformal invariance of $\langle \text{F}|\text{G}\rangle$ in \Eq{eq:scalarproduct} is crucial for a consistent interpretation of measurements \cite{Sorkin1994,Hardy2001} which, besides quantum mechanics, can also be used in wave mechanics. In this interpretation, an {\em observable} property is represented by a self--adjoint operator $\Gamma$, and the value of the property in a ket is computed by the trace rule, which for a pure state such as $|\text{F}\rangle$ leads to the ``sandwiches'' in \Eq{eq:gamma} that we consider in this paper
	\begin{equation}
		\label{eq:tracerule}
		\text{Trace}\{\Gamma |\text{F}\rangle\langle\text{F}|\}= \langle\text{F}|\Gamma|\text{F}\rangle.
	\end{equation}

	The fact that \Eq{eq:tracerule} is useful for Maxwell fields is clearly seen in e.g. \cite{Birula1996}. For example, it is there shown that the result of $\langle\Phi|\text{H}|\Phi\rangle$ for the energy operator $\text{H}$, and of $\langle\Phi|\PP|\Phi\rangle$ for the momentum operator vector $\mathbf{P}$ are equivalent to the corresponding integrals in \Eq{eq:hpdr}. Besides reproducing results typically obtained by other means, the algebraic approach for new ones to be derived, such as the content of the generators of Lorentz boosts in a given field \cite[Equation~(4.16)]{Birula1996}, as well as for an alternative expression for the optical helicity \cite{FerCor2019b}.

\begin{figure}[t!]
	\includegraphics[width=\linewidth]{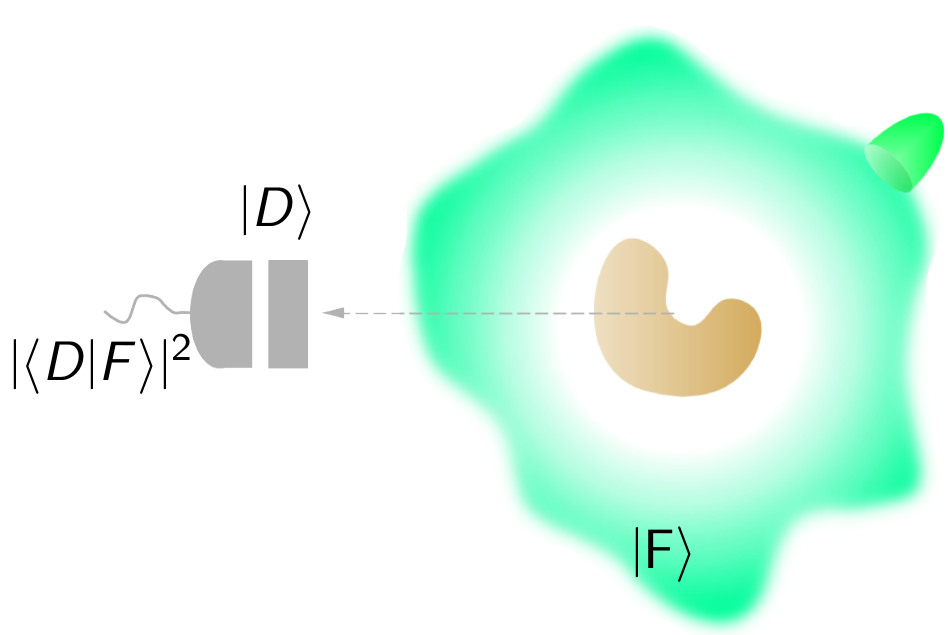}
		\caption{\label{fig:measurement} In a projective measurement, the outcome of an apparatus for measuring the field $|\text{F}\rangle$ can be modeled as $|\langle \text{D}|\text{F}\rangle|^2$, that is, the modulus square of the projection of $|\text{F}\rangle$ onto an electromagnetic mode $|\text{D}\rangle$.}
\end{figure}

		In order to better appreciate the importance of an invariant scalar product, let us now examine its role in the consistent definition of projective measurements. Consider a setup such as the one in Fig.~\ref{fig:measurement}, where the field $|\text{F}\rangle$ is measured. We can imagine, for example, that $|\text{F}\rangle$ is the outgoing field in the right hand side of Fig.~\ref{fig:lmi}, although the exact way in which the field is produced is not important. At a certain point far away from the material object, that is, in the far field, we place a measurement device consisting of an analyzer that selects a particular frequency $\omega_0$ and polarization $\sigma$, followed by a photo--detector. The number of clicks in the detector will be equal to $|\langle \text{D}|\text{F}\rangle|^2$, where $|\text{D}\rangle$ is essentially a $\sigma$--polarized plane wave with momentum $\omega_0/c_0\mathbf{\hat{d}}$ lying in the direction connecting the origin of coordinates (assumed to be in or nearby the object) and the location of the measurement device. 

Let us now consider the effects of a global transformation X applied to the whole physical system including both the field and the measurement apparatus:
	\begin{equation}
		\label{eq:Xtrans}
			|\text{F}\rangle \rightarrow\text{X}|\text{F}\rangle,\  |\text{D}\rangle\rightarrow\text{X}|\text{D}\rangle \implies |\langle \text{D}|\text{F}\rangle|^2\rightarrow|\langle \text{D}|\text{X}^\dagger\text{X}|\text{F}\rangle|^2.
	\end{equation}

	In order for the measurement outcome $|\langle \text{D}|\text{F}\rangle|^2$ to be meaningful, we must require that $|\langle \text{D}|\text{F}\rangle|^2=|\langle \text{F}|\text{X}^\dagger\text{X}|\text{D}\rangle|^2$ for any X in the largest symmetry group of Maxwell equations, i.e., the conformal group $C_{15}(3,1)$. Measurement outcomes should not change under the allowed changes of reference frame or changes of units. Similarly, we must also require that quantities such as $\langle \text{F}|\Gamma|\text{F}\rangle$, which can be interpreted as the average value of the measurements of $\Gamma$ on $|\text{F}\rangle$, are also invariant under any X in the conformal group: $\langle\text{F}|\Gamma|\text{F}\rangle=\langle\text{F}|\text{X}^\dagger\text{X}\Gamma\text{X}^{-1}\text{X}|\text{F}\rangle$. 

	 All such invariance requirements are fulfilled because any X in the conformal group is unitary with respect to the scalar product in \Eq{eq:scalarproduct} \cite{Gross1964}, and hence $\text{X}^\dagger\text{X}$ is the identity.

	 Thus, it is clear that the extension of the algebraic approach to matter requires a group of transformations and an invariant scalar product for the static case.

 	\section{Conformally invariant scalar product for static matter\label{sec:sp}}
	\subsection{Representation of matter}
	With respect to electromagnetism, matter in static equilibrium can be represented by its electric charge density $\rhorest$ and its magnetization density $\mrest$. We assume that the time derivatives of macroscopic quantities vanish, and that there are no static currents [$\mathbf{J}(\rr)=\zerovec$]. We also assume that $\rhor$ and $\Mr$ are contained in a finite volume. 

The choice of $\rhor$ and $\Mr$ for representing matter in static equilibrium is motivated by the existence of electric charge and magnetic spin as fundamental properties of elementary particles. In sharp contrast, while the search continues \cite{Rajantie2016,Fortson2003}, there is no experimental evidence of isolated magnetic charges or static electric dipole moments. We therefore exclude magnetic charge densities $\rho_{\text{mag}}(\rr)$ and static polarization densities $\mathbf{P}(\rr)$ from the description of matter in static equilibrium. We note that the static electric dipoles that are present in certain molecules can be described by the dipole moment of their electric charge density, and that the {\em dynamic} polarization density $\Prt$ can be understood as arising from a nonstatic magnetization density. In such model $\Mrt$ and $\Prt$ are the space--space and time--space components of a totally antisymmetric tensor, respectively, and $\Prt$ vanishes in static equilibrium ($\mathbf{P}(\rr)=\zerovec$). Such a totally antisymmetric tensor is a common model for point particles in relativistic electrodynamics \cite[Chapter~II, Section~4]{Barut1980}, and has been used in studies of the effect of the electron spin on the atomic nucleus \cite{Frenkel1926}, and of the relativistic spin precession \cite{Bargmann1959}.

	The densities $\rhorest$ and $\mrest$ are equivalent to the static fields that they generate, as per the equations of electrostatics and magnetostatics. The scalar charge density generates the Coulomb field, which is a longitudinal (zero-curl) electric field:
\begin{equation}
	\label{eq:Er}
	\epsz\nabla\cdot\Er=\rhor,\ \nabla\times\Er=\zerovec.
\end{equation}

	The vectorial magnetization density generates the $\Br$ and $\Hr$ fields. Outside the material object, both fields are proportional to each other and transverse (i.e., have zero divergence). Inside the object, where $\Mr\neq \zerovec$, the $\Br$ field is transverse, and the $\Hr$ field is longitudinal:
\begin{equation}
	\label{eq:staticeqsr}
	\begin{split}
		\Br/\muz-\Hr=\Mr,&\ \nabla\cdot\Hr=-\nabla\cdot\Mr,\\ 
	\nabla\times\Br/\muz=\nabla\times\Mr&,\ \nabla\cdot\Br=0.
	\end{split}
\end{equation}

Equation~(\ref{eq:Er}) and the second line of \Eq{eq:staticeqsr} can be obtained as the limit of Maxwell equations in static equilibrium \cite[Section~III]{FerCor2020}, and the first line of \Eq{eq:staticeqsr} is the definition of $\Hr$. The expressions in \Eq{eq:staticeqsr} can alternatively be obtained by imposing our assumption $\mathbf{J}(\rr)=\zerovec$ onto the typical magnetostatic equations found in e.g. \cite[Equations~(2.37),(2.40),(2.41)]{Brown1962} or in \cite[Equations~(5.80)-(5.82)]{Jackson1998}. Effectively, the equations in (\ref{eq:staticeqsr}) are identifications of the transverse and longitudinal parts of $\Mr$ with the other fields, and we may as well use only $\Mr$ instead of both $\Br$ and $\Hr$ together. Equation~(\ref{eq:Er}) implies that $\Er$ and $\rhor$ determine each other bijectively. We choose to use $\Er$ here because it shares some transformation properties with the magnetization $\Mr$, allowing us to avoid different derivations in some cases, should we use $\rhor$ instead. We can therefore represent matter in static equilibrium by means of $\Er$ and $\Mr$, or alternatively by means of their Fourier transforms $\mathbf{E}(\pp)$ and $\mathbf{M}(\pp)$, which can be obtained by integrals in the finite volume V occupied by the object:
	\begin{equation}
		\label{eq:epmp}
		\begin{split}
			\mathbf{M}(\pp)&=\intdpnormV \mathbf{M}(\rr)\exp(-\ii\pp\cdot\rr),\\
			\mathbf{E}(\pp)&=\frac{-\ii \phat}{\epsz|\pp|}\rho(\pp)=\frac{-\ii \phat}{\epsz|\pp|}\intdpnormV \rho(\rr)\exp(-\ii\pp\cdot\rr),
		\end{split}
	\end{equation}
	where the first equality in the second line of \Eq{eq:epmp} follows from \Eq{eq:Er}.

	We note that both $\Er$ and $\Mr$ are real--valued.

\subsection{Group of transformations and invariant scalar product for the static case}
	In this section, we define an appropriate scalar product for static matter, which, in particular, allows one to use \Eq{eq:gamma} for computing the amount of fundamental quantities stored in matter. With such scalar product, each $\{\rhor,\Mr\}$ corresponds to a ket $\Phistaticket$ in the Hilbert space of static matter. This Hilbert space is different from the one containing the radiation fields. One salient difference is the frequency $\omega$, which is equal to zero for static matter and strictly larger than zero in radiation fields. Another difference is that there are helicity zero (longitudinal) components in static matter, while the radiation fields are always transverse.

	Motivated by the requirement of invariant measurements formulated in Section~\ref{sec:measurement}, we pursue the idea of a group invariance for the scalar product by starting with the full conformal invariance of Maxwell equations with sources shown by Bateman \cite{Bateman1910}, Cunningham \cite{Cunningham1910} and Dirac \cite{Dirac1936}, among others \cite{Kastrup2008}. While in Dirac's work the sources were electric charge--current densities, both Bateman and Cunningham additionally included magnetization and polarization densities. 

	The conformal invariance including sources deserves further discussion. On the one hand, the invariance of Maxwell equations with sources under the 15 parameter conformal group in Minkowski spacetime $C_{15}(3,1)$ is often explicitly recognized in the literature; e.g., see \cite[Section~3.1]{Fuschchich1994}, or the review by Kastrup \cite{Kastrup2008} where the conformal symmetry, its role in theoretical physics, and their historical evolution are comprehensively explained. On the other hand, the fact that fixed mass parameters break the scale invariance, and consequently the conformal invariance, could suggest the incorrect conclusion that Maxwell equations are conformally invariant only in the source--free case. In this respect, it is important to note that Maxwell equations including electric charge--current densities $[\rhort,\Jrt]$, and magnetization and polarization densities $[\Mrt,\Prt]$ {\em do not feature any fixed mass parameter}. The invariance under conformal transformations follows from the way in which the sources transform. The transformations of $[\rhort,\Jrt]$ can be found in \cite[Equations~3.40ab]{Fuschchich1994}. A simple illustration is the transformation law of the electric charge density under a scaling $\rr\rightarrow \alpha\rr$: $\rho(\rr)\rightarrow \rho(\rr/\alpha)/\alpha^3$ in the static case, which is readily shown to preserve the total charge in a volume. The bottom line is that when both fields and sources are transformed, the form of the dynamic Maxwell equations {\em with sources} remains invariant. This is what we use below as the starting point for obtaining the sought after expression of a scalar product for the static case. Incidentally, even equations with a mass parameter, such as the Dirac equation, can be shown to be conformally invariant if one allows for a particular re--scaling of the mass \cite[Section~5.1]{Kastrup2008}. 

Let us now return to the idea of invariance for the scalar product. Since we are considering matter in static equilibrium, we need to remove all the transformations of $C_{15}(3,1)$ that do not preserve the $\omega=0$ condition. Lorentz boosts change $\omega$ and mix it with the components of $\pp$. The time component of the special conformal transformations four-vector also changes $\omega$ \cite[Section~3]{Budinich1993}. We remove all these transformations. We also remove time--translations because, while preserving the $\omega=0$ condition, any time--translations will, for the static case, just degenerate into the identity operator. We are then left with a ten parameter group consisting of spatial translations, spatial rotations, the spatial scaling ($\rr\rightarrow\alpha\rr$ with $\alpha\in\mathbb{R},\ \alpha>0$~), and three special conformal transformations: The ten parameter conformal group $C_{10}(3)$ in three-dimensional Euclidean space \cite[Chapter~24]{Grafarend2014}. It is interesting to see that the static restriction of the conformal group in 3+1 spacetime dimensions results in the conformal group in three spatial dimensions. Accordingly, along with the discussion in Section~\ref{sec:measurement}, our sought--after scalar product must be invariant under all the transformations in $C_{10}(3)$. 

	We now proceed by considering the $C_{15}(3,1)$-invariant scalar product expression in \Eq{eq:scalarproduct}. We first perform a unitary change of basis, going from the $\mathbf{F}_{\pm}(\pp)$ to the $\{\sqrt{\epsilon_0}\mathbf{E}(\pp),\ii\mathbf{B}(\pp)/\sqrt{\mu_0}\}$ basis by using \Eq{eq:frt} to express the latter as linear combinations of the former. 
	\begin{equation}
		\label{eq:ai}
		\frac{1}{\sqrt{2}}\begin{bmatrix}\Id&\Id\\-\ii\Id&\ii\Id\end{bmatrix}\begin{bmatrix}\Fpplus\\\Fpminus\end{bmatrix}=\begin{bmatrix}\sqrt{\epsz}\ \mathbf{E}(\pp)\\\mathbf{B}(\pp)/\sqrt{\muz}\end{bmatrix},
	\end{equation}
	where $\Id$ is the 3$\times$3 identity matrix. We now replace $\mathbf{B}(\pp)$ by $\muz\Mp$. Both quantities have the same units. This replacement is necessary in order to include the longitudinal degree of freedom that $\Mp$ can contain, which is always absent in $\mathbf{B}(\pp)$: 
	\begin{equation}
		\label{eq:aiai}
		\begin{bmatrix}\sqrt{\epsz}\ \mathbf{E}(\pp)\\\mathbf{B}(\pp)/\sqrt{\muz}\end{bmatrix}\rightarrow\begin{bmatrix}\sqrt{\epsz}\ \mathbf{E}(\pp)\\\sqrt{\muz}\Mp\end{bmatrix}.
	\end{equation}

	Equations (\ref{eq:scalarproduct}), (\ref{eq:ai}), and  (\ref{eq:aiai}) motivate us to write the following scalar product expression for the $\omega=0$ fields:

\begin{empheq}[box=\fbox]{align}
    &\langle\Phi_{\omega=0}^1|\Phi_{\omega=0}^2\rangle=\nonumber\\ 
	& \intdpconfhbar \begin{bmatrix}\sqrt{\epsz}\ \mathbf{E}^1(\pp)\\\sqrt{\muz}\ \mathbf{M}^1(\pp)\end{bmatrix}^\dagger \begin{bmatrix}\sqrt{\epsz}\ \mathbf{E}^2(\pp)\\\sqrt{\muz}\ \mathbf{M}^2(\pp)\end{bmatrix}.
\label{eq:static}
   \end{empheq}

As is the case in \Eq{eq:scalarproduct}, the quantity under the integral sign in \Eq{eq:static} is unitless. This is readily seen with the following equalities between the units of different fields:
\begin{equation}
	\nonumber
		\left[\Fplambda\right]\equalduetoeq{eq:frt}\left[\sqrt{\epsz}\mathbf{E}(\pp)\right]=\left[\sqrt{\epsz}\cz\mathbf{B}(\pp)\right]\equalduetoeq{eq:staticeqsr}\left[\sqrt{\muz}\Mp\right].
\end{equation}
.

When written in the equivalent form in \Eq{eq:fsp}, \Eq{eq:static} coincides with the invariant scalar product for the relevant representations of the scale--Euclidean group in \cite[Equation~(29)]{Moses1973}. The extra $1/|\pp|^2$ in Equation~(29) of \cite{Moses1973} with respect to \Eq{eq:fsp} in this paper is compensated by the extra $1/|\pp|$ factor in the plane wave decomposition in \cite[Equation~(57)]{Moses1973} as compared to \Eq{eq:epmp}. That $N=-2$ in \cite[Equation~(57)]{Moses1973} follows from the definition of $N$ in \cite[Equation~(25a)]{Moses1973}, and the transformation properties of $\Er$ and $\Mr$ under spatial scalings. Therefore, the scalar product in \Eq{eq:static} is invariant under the scale--Euclidean group, which is a seven--parameter sub--group of $C_{10}(3)$ composed by translations, rotations, and spatial scalings. These transformations act hence unitarily with respect to the scalar product. The invariance of the scalar product in \Eq{eq:static} under the rest of $C_{10}(3)$, that is, under special conformal transformations, can be seen as follows.

The special conformal transformations act on the coordinate vector as \cite[Equation~(24.4)]{Grafarend2014}:
\begin{equation}
	\rr \rightarrow \frac{\rr+\mathbf{c}|\rr|^2}{1+2\mathbf{c}\cdot\rr+|\mathbf{c}|^2|\rr|^2},
\end{equation}
where $\mathbf{c}$ is a real--valued 3-vector: $c_l \in \mathbb{R},\ l=1,2,3$. To prove that \Eq{eq:static} is invariant under such transformations one must show that the special conformal transformations are unitary. This unitary character is shown by considering that the special conformal transformation $\mathrm{C}_l(c_l)$ along axis $l$ with parameter $c_l$ can be obtained from the spatial translation along such axis by the same parameter $\mathrm{T}_l(c_l)$, and the inversion operation $\RR$ \cite[below Equation~(24.4)]{Grafarend2014}:
\begin{equation}
	\label{eq:rtr}
	\mathrm{C}_l(c_l)=\RR \mathrm{T}_l(c_l) \RR\text{, for }l=1,2,3.
\end{equation}

The action of $\RR$ on the coordinate vector is
\begin{equation}
	\rr \rightarrow \frac{\rr}{|\rr|^2}.
\end{equation}

Since we already known that the translations act unitarily, \Eq{eq:rtr} implies that and if $\RR$ is unitary, then $C_l$ is unitary as well. The unitary character of $\RR$ is shown in App.~\ref{app:runitary} by extending a result contained in Section~II of \cite{Kastrup1970}. 

Once $\RR$ is known to be unitary, the conclusion that the scalar product in \Eq{eq:static} is invariant under $C_{10}(3)$ can also be reached directly without using its invariance under the scale--Euclidean group. First, we establish that the action of spatial translations leave \Eq{eq:static} invariant, which is obvious when substituting the known action of arbitrary translations by a displacement vector $\mathbf{c}$ on the Fourier transforms  $\mathbf{E}(\pp)$ and $\mathbf{M}(\pp)$
\begin{equation}
	\mathbf{E}(\pp)\rightarrow \exp(-\ii \mathbf{c}\cdot\pp)\mathbf{E}(\pp),\ \mathbf{M}(\pp)\rightarrow \exp(-\ii \mathbf{c}\cdot\pp)\mathbf{M}(\pp),
\end{equation}
onto \Eq{eq:static}, where the acquired phases cancel.

Since the translations are unitary, their generators $\mathrm{P}_l$ are self--adjoint, and so are the generators of the special conformal transformations $\mathrm{K}_l$, because of \Eq{eq:rtr}. The other generators of $C_{10}(3)$, that is, the angular momentum operators and the generator of dilations $\text{D}$, can be obtained as commutators of $\mathrm{K}_l$ and $\mathrm{P}_i$ \cite[Equation~(20)]{Kastrup1970}:
\begin{equation}
	\label{eq:all}
	[\mathrm{K}_l,\mathrm{P}_i]=2\ii\left(\delta_{li}\mathrm{D}-\mathrm{M}_{li}\right)\text{, where } l,i=1,2,3.
\end{equation}

Then, because the adjoint of the commutator of two self--adjoint operators is minus itself, it follows, noting that $\ii\rightarrow -\ii$ when taking the adjoint of the right hand side of \Eq{eq:all}, that $\mathrm{D}$ and $\mathrm{M}_{li}$ must be self--adjoint, and hence the transformations that they generate must be unitary. We recall that $\mathrm{M}_{ll}=0$, $\mathrm{M}_{li}=-\mathrm{M}_{il}$, $\mathrm{M}_{12}=\mathrm{J}_3$, $\mathrm{M}_{23}=\mathrm{J}_1$, and $\mathrm{M}_{31}=\mathrm{J}_2$.

We adopt \Eq{eq:static} as the $C_{10}(3)$-invariant scalar product for $\omega=0$. This provides a new way of obtaining expressions for computing the amount of fundamental quantities stored in matter.

Conveniently, the very similar functional form of the expressions for the scalar products in the dynamic and static cases, Equations~(\ref{eq:scalarproduct}) and (\ref{eq:static}) respectively, implies that known $\pp$--space expressions of fundamental operators for the $\omega>0$ case \cite{Moses1965,Birula1975,Birula1996} can be also used for the $\omega=0$ case. The same is true for the corresponding $\rr$--space expressions. 

We note that, while the scalar products for the dynamic case in \Eq{eq:scalarproduct} and for the static case in \Eq{eq:static} are invariant under $C_{15}(3,1)$ and $C_{10}(3)$, respectively, the dynamic fields and static densities are not required to exhibit specific symmetries.

In the rest of the article, we will focus on fundamental quantities stored in the magnetization density, for which we set $\Er=\zerovec$ in \Eq{eq:static}:
	\begin{equation}
		\label{eq:mronly}
		\langle\Phi_{\omega=0}^1|\Phi_{\omega=0}^2\rangle=\langle \mathrm{M}_1|\mathrm{M}_2\rangle = \intdpM \left[\mathbf{M}^1(\pp)\right]^\dagger\mathbf{M}^2(\pp).
	\end{equation}

Nevertheless, the methodology that we use applies to the case with $\Er\neq\zerovec$ as well. For example, the representations of $\Mr$ using three scalar complex functions in the domain of linear momentum  $\ff{\lambda=-1,0,1}$, or angular momentum $\mathrm{f}_{jm\left(\lambda=-1,0,1\right)}(|\pp|)$, where $\lambda$ is the helicity, have their counterparts for representing $\Er$, except that $\Er$ contains only the $\lambda=0$ component. The counterpart of \Eq{eq:f} is
\begin{equation}
	\ff{0}={\mathbf{e}_0(\phat)}^\dagger\sqrt{\frac{\epsz}{\cz\hbar}}\mathbf{E}(\pp),
\end{equation}
and Equations~(\ref{eq:ffromF}),(\ref{eq:Fsp}),(\ref{eq:am}), and (\ref{eq:jzjpjm}) apply, with the difference that the helicity takes only the value $\lambda=0$. 

\section{Helicity and angular momentum in magnetization\label{sec:form}}
	We start with the helicity stored in $\Mr$. At each $\pp$ point, $\Mp$ can be decomposed into the three eigenstates of the helicity operator for vectorial fields:
\begin{equation}
	\label{eq:Mlambda}
	\Mp = \sum_{\lambda=-1,0,1} \mathbf{M}_\lambda(\pp),\ \hbar\Help \mathbf{M}_\lambda(\pp) = \hbar\lambda \mathbf{M}_\lambda(\pp).
\end{equation}
The longitudinal $\lambda=0$ component corresponds to non--vanishing $\nabla\cdot\Mr$. While a net magnetic charge in an isolated object has not been observed, complicated magnetization textures are expected to contain pairs of monopoles of opposite charge \cite{Liu2018}, and pairs of singularities have indeed been experimentally imaged \cite{Donnelly2017}. The longitudinal $\lambda=0$ term in the magnetization is crucial for their description. In contrast, the free dynamic electromagnetic field is divergenceless, and the $\lambda=0$ component vanishes.

For the explicit decomposition into plane waves we use 
\begin{equation}
	\label{eq:mq}
	\sqrt{\frac{\muz}{\cz\hbar}}\mathbf{M}_\lambda(\pp)= \ff{\lambda}\mathbf{e}_\lambda(\phat),
\end{equation}
where $\ff{\lambda}$ are complex--valued functions, and 
\begin{equation}
	\label{eq:epmez}
		\mathbf{e}_0(\phat)=\frac{\pp}{|\pp|}=\phat,\ \mathbf{e}_{\pm}(\phat)=\frac{1}{\sqrt{2}}\begin{bmatrix}\mp \cos\phi\cos\theta+\ii\sin\phi\\\mp \sin\phi\cos\theta-\ii\cos\phi\\\pm \sin\theta\end{bmatrix},
\end{equation}
where $\phi=\atantwo(k_y,k_x)$, and $\theta=\arccos(k_z/|\pp|)$. The $\mathbf{e}_\lambda(\phat)$ meet \cite{Birula1975,Birula1996}

\begin{equation}
	\label{eq:qprops}
	\Help \mathbf{e}_\lambda(\phat) = \lambda \mathbf{e}_\lambda(\phat)\text{, and }\left[\mathbf{e}_\lambda(\phat)\right]^\dagger\mathbf{e}_{\bar{\lambda}}(\phat)=\delta_{\lambda\bar{\lambda}}.
\end{equation}

The last equation in (\ref{eq:qprops}) implies that
\begin{equation}
	\label{eq:f}
	\ff{\lambda}={\mathbf{e}_\lambda(\phat)}^\dagger \sqrt{\frac{\muz}{\cz\hbar}}\Mp.
\end{equation}

Other choices for the helicity eigenvectors exist in the literature. In particular, the $\mathbf{e}_\lambda(\phat)$ used by Bialynicki--Birula \cite{Birula1975,Birula1996} are related to the $\Qlambda$ vectors used by Moses \cite{Moses1973,Moses2004} in the following way: \mbox{$\Qlambda = -\mathbf{e}_\lambda(\phat)\exp(i\lambda\phi)$}.

Using the expression of the helicity operator in \Eq{eq:Hel}, the helicity stored in $\Mr$ can then be written as:
\begin{equation}
	\label{eq:helf}
	\begin{split}
		&	\langle \text{M}|\Lambda|\text{M}\rangle =\intdpM\left[\Mp\right]^\dagger\hbar\Help \Mp\\
		&\equaldueto{\text{Equation~(\ref{eq:Mlambda})}}\intdpM\left[\sum_{\bar{\lambda}=-1,0,1}\mathbf{M}_{\bar{\lambda}}(\pp)\right]^\dagger\sum_{\lambda=-1,0,1}\hbar\lambda \mathbf{M}_\lambda(\pp)\\
		&\equaldueto{\text{Equations~(\ref{eq:mq},\ref{eq:qprops})}}\hbar\intdpMsp\left[ |\ff{+}|^2-|\ff{-}|^2\right].
	\end{split}
\end{equation}

We note that the charge density cannot store helicity, because $\Help \mathbf{E}(\pp)=\Help \frac{-\ii\phat\rho(\pp)}{\epsz|\pp|}=\zerovec$. 

As shown in \cite{FerCor2020}, the result of \Eq{eq:helf} is proportional to the definition of the static magnetic helicity \cite{Woltjer1958,Moffatt1969,Ranada1992}:
\begin{equation}
	\int_{\mathbb{R}^3} \mathrm{d}^3 \rr \text{ } \Br\cdot\Ar = \int_{\mathbb{R}^3} \mathrm{d}^3 \pp \ \mathbf{B}^\dagger(\pp)\mathbf{A}(\pp),
\end{equation}
where $\nabla\times \Ar =\Br$, and hence $\ii\pp\times \mathbf{A}(\pp)= \mathbf{B}(\pp)$.

More generally, Equations~(\ref{eq:Mlambda}), (\ref{eq:qprops}), and (\ref{eq:f}) can be used to write the expression of the scalar product in \Eq{eq:mronly}
 as a function of the $\ff{\lambda}$
	\begin{equation}
		\label{eq:fsp}
		 \langle \mathrm{M}_1|\mathrm{M}_2\rangle =\sum_{\lambda=-1,0,+1}\intdpMsp \left[\prescript{1}{}{\ff{\lambda}}\right]^*\prescript{2}{}{\ff{\lambda}},
	\end{equation}
	which allows one to compute $\langle \mathrm{M}|\Gamma|\mathrm{M}\rangle$ from the $\ff{\lambda}$, as long as the action of $\Gamma$ on $\ff{\lambda}$ is known.

	Let us now turn our attention to angular momentum. Rather than using $\ff{\lambda}$, the computation of $\langle \text{M}|\text{J}_i|\text{M}\rangle$ is more conveniently performed in the angular momentum basis:
\begin{equation}
	\label{eq:mp}
	\Mr\equiv \sum_{\lambda={-1,0,1}}\sum_{j=1}^\infty\sum_{m=-j}^j\int_{0}^\infty \text{d}|\pp| \FF{\lambda} |\pp||k\ j\ m\ \lambda\rangle,
\end{equation}
where $|k\ j\ m\ \lambda\rangle$ denotes a simultaneous eigenstate of \protect{$\PP\cdot\PP=\text{P}^2$}, \protect{$\JJ\cdot\JJ=\text{J}^2$}, $\text{J}_z$, and $\Lambda$, with eigenvalues $\hbar^2|\pp|^2=\hbar^2 k^2$, $\hbar^2 j(j+1)$, $\hbar m$, and $\hbar \lambda$, respectively, and 
	\begin{equation}
		\label{eq:ffromF}
		\FF{\lambda}=\int \text{d}^2\phat \ \sqrt{\frac{2j+1}{4\pi}} \text{D}_{m\lambda}^j(\phat)\ff{\lambda},
	\end{equation}
	where the Wigner D--matrices for spatial rotations \cite[Chapter~4]{Varshalovich1988} enter as follows
	\begin{equation}
		\label{eq:Dexplicit}
		\text{D}_{m\lambda}^j(\phat)=\text{D}_{m\lambda}^j(\phi,\theta,0)=\exp(-\ii m\phi)\text{d}^j_{m\lambda}(\theta),
	\end{equation}
	where $\text{d}_{m\lambda}^j(\theta)$ are the Wigner small d--matrices as defined in \cite[Chapter~4.3]{Varshalovich1988}, and $\int \text{d}^2\phat\equiv\int_{0}^\pi \text{d}\theta\sin\theta\  \int_{-\pi}^\pi \text{d}\phi$. The publicly available EasySpin computer code contains a convenient implementation of the Wigner matrices \cite{Stoll2006}.

	As we show in App.~\ref{sec:spam}, the scalar product $\langle \mathrm{M}_1|\mathrm{M}_2\rangle$ can also be written as:
	\begin{equation}
		\label{eq:Fsp}
		\begin{split}
			&\langle \mathrm{M}_1|\mathrm{M}_2\rangle=\\
			&\sum_{\lambda=-1,0,+1}\sum_{j=1}^\infty\sum_{m=-j}^{j}\intdmodpM\left[\prescript{1}{}{\FF{\lambda}}\right]^*\prescript{2}{}{\FF{\lambda}}.
		\end{split}
	\end{equation}
The action of angular momenta operators on $\FF{\lambda}$ is \cite[Equations~(2.1)-(2.3)]{Lomont1964}:
\begin{equation}
	\label{eq:am}
	\begin{split}
		&\frac{\text{J}_z}{\hbar}\FF{\lambda}= m\FF{\lambda},\\
		&\frac{\left(\text{J}_y+\ii\text{J}_x\right)}{\hbar}\FF{\lambda}=\sqrt{(j-m)(j+m+1)}\mathrm{f}_{j(m+1)\lambda}(|\pp|),\\
		&\frac{\left(\text{J}_y-\ii\text{J}_x\right)}{\hbar}\FF{\lambda}=\sqrt{(j+m)(j-m+1)}\mathrm{f}_{j(m-1)\lambda}(|\pp|),\\
		&\frac{\text{J}^2}{\hbar^2}\FF{\lambda}=\frac{\left(\text{J}_x^2+\text{J}_y^2+\text{J}_z^2\right)}{\hbar^2}\FF{\lambda}=j(j+1)\FF{\lambda},
	\end{split}
\end{equation}

with which we can write: {\small
\begin{equation}
	\label{eq:jzjpjm}
	\begin{split}
		&\langle \mathrm{M}|\text{J}_z|\mathrm{M}\rangle=\hbar\sum_{\lambda=-1,0,+1}\sum_{j=1}^\infty\sum_{m=-j}^{j}\intdmodpM m|\FF{\lambda}|^2,\\
		&\langle \mathrm{M}|\text{J}_y+\ii\text{J}_x|\mathrm{M}\rangle=\\&\hbar\sum_{\lambda j m}\intdmodpM \sqrt{(j-m)(j+m+1)}\left[\FF{\lambda}\right]^*\mathrm{f}_{j(m+1)\lambda}(|\pp|),\\
		&\langle \mathrm{M}|\text{J}_y-\ii\text{J}_x|\mathrm{M}\rangle=\\&\hbar\sum_{\lambda j m}\intdmodpM\sqrt{(j+m)(j-m+1)} \left[\FF{\lambda}\right]^*\mathrm{f}_{j(m-1)\lambda}(|\pp|),\\
		&\langle \mathrm{M}|\text{J}^2|\mathrm{M}\rangle=\hbar^2 \sum_{\lambda j m}\intdmodpM j(j+1)|\FF{\lambda}|^2.\\
	\end{split}
\end{equation}
}

Appendix~\ref{sec:iszero} shows that the angular momenta $\langle \mathrm{M}|\text{J}_{\tau \in \{x,y,z\}}|\mathrm{M}\rangle$ and the linear momenta $\langle \mathrm{M}|\text{P}_{\tau\in \{x,y,z\}}|\mathrm{M}\rangle$ both vanish. The origin of this vanishing can be found in the following restrictions, which follow from assuming $\mrest\in\mathbb{R}^3$:  
\begin{equation}
	\label{eq:real}
	\begin{split}
		\ff{\lambda}&=(-1)^{\lambda+1}\left[\text{f}_{\lambda}(-\pp)\right]^*,\\
		\FF{\lambda}&=(-1)^{j-m+1}\left[\text{f}_{j(-m)\lambda}(|\pp|)\right]^*.
	\end{split}
\end{equation}

For example, for $\langle \mathrm{M}|\text{J}_z|\mathrm{M}\rangle$, the proof is straightforward because the sum over $m$ vanishes since the second line of Equation~(\ref{eq:real}) implies that $|\mathrm{f}_{jm\lambda}(|\pp|)|^2=|\mathrm{f}_{j(-m)\lambda}(|\pp|)|^2$, and hence $m|\mathrm{f}_{jm\lambda}(|\pp|)|^2-m|\mathrm{f}_{j(-m)\lambda}(|\pp|)|^2=0$ for all $|m|\in[0,j]$. 

Derivations similar to those in App.~\ref{sec:iszero} show that the linear and angular momenta in $\rhorest$[$\Er$] vanish as well since, for $\Er\in\mathbb{R}^3$: 
\begin{equation}
	\label{eq:realE}
	\begin{split}
		\ff{0}&=-\left[\text{f}_{0}(-\pp)\right]^*,\\
		\FF{0}&=(-1)^{j-m+1}\left[\text{f}_{j(-m)0}(|\pp|)\right]^*.
	\end{split}
\end{equation}

The vanishing of linear and angular momenta for the whole $\Phistaticket$ is consistent with the static equilibrium condition under zero external fields. In contrast, the squared linear and angular momenta $\langle \mathrm{M}|\text{P}^2|\mathrm{M}\rangle$, and $\langle \mathrm{M}|\text{J}^2|\mathrm{M}\rangle$, do not vanish [see e.g. the last line of \Eq{eq:jzjpjm}]. We note that the non-zero mean value of the square of a property simultaneously with zero mean value of such property can easily occur, and should not be taken as a sign of a quantum effect. Since the eigenvalues of self--adjoint operators are real numbers, the eigenvalues of squares of self--adjoint operators are non--negative real numbers. Then, the value of a property measured as the average of a self--adjoint operator for a given state can be zero, while, simultaneously, the average of the squared operator is non--zero. For example: The average joint linear momentum of two counterpropagating plane--waves is zero in any direction, yet the linear momentum squared ($\PP\cdot\PP$) is not.

The $\ff{\lambda}$ and $\FF{\lambda}$ for a given $\Mr$ can be readily obtained with the following sequence of computations,
\begin{equation}
	\label{eq:sequence}
	\Mr\stackrel{\Eq{eq:epmp}}{\rightarrow}\Mp\stackrel{\Eq{eq:f}}{\rightarrow}\ff{\lambda}\stackrel{\Eq{eq:ffromF}}{\rightarrow}\FF{\lambda}.
\end{equation}
which can be applied to analytically derived, numerically obtained, or experimentally measured three-dimensional magnetization textures confined to finite volumes.

We note that, given $\rhor$, the $\ff{0}$ and $\FF{0}$ corresponding to $\mathbf{E}(\pp)$ can be obtained similarly.

\subsection{Discrepancies with existing results\label{sec:diff}}

The computation of the linear and angular momenta of magnetic solitons \cite{Dzyloshinskii1979,Volovik1987,Papanicolaou1991,Papanicolaou1993,Cooper1999,Kosevich1990} remains an active field of research \cite{Borisov2008,Borisov2009,Yan2013,Tchernyshyov2015}. Starting from the Landau--Liftshitz (LL) equation, different integral expressions have been proposed, mostly for predicting the dynamics of topologically non--trivial magnetization textures inside an unbounded magnetic medium under, for example, an external magnetic field. Nevertheless, such integrals are assumed to also be valid for the static case without external fields. Using one of the latest versions of such integrals, even a completely static domain wall has a finite linear momentum \cite{Yan2013}. With our methodology, as shown in App~\ref{sec:piszero}, a static magnetization density has a vanishing value of average linear momentum and angular momentum along any given axis. The difference in the results can be attributed to the fact that the two approaches are applied to systems meeting quite different assumptions. In this article, we consider a bounded three-dimensional domain containing a given static three-dimensional magnetization. In contrast, the LL--based approaches consider a two or three-dimensional domain of infinite size where the magnetization is allowed to vary, as for example when three-dimensional topological textures are stabilized by endless motion with constant velocity \cite{Papanicolaou1993}.

\subsection{Helicity and angular momentum of a Hopfion\label{sec:hopfion}}
As an exemplary application, we will now compute the helicity and angular momentum squared stored in a Hopfion.

A convenient analytical approximation of the magnetization density of a Hopfion has been provided by \mbox{P.~Sutcliffe} in \cite{Sutcliffe2018}, where the Hopfion is hosted in a disk of FeGe of height L=\SI{70}{\nano\meter} and diameter 3L, and is numerically stabilized to a cylindrically symmetric configuration using the Dzyaloshinskii--Moriya interaction of the chiral magnet and a strong anisotropy perpendicular to the flat ends of the cylinder. With good approximation to the numerical results, the Cartesian components of unit magnetization density vector $\mhatr$ of the Hopfion are given in \cite[Equation~(3.3)]{Sutcliffe2018} as a function of the cylindrical coordinates of the position vector $\rr\equiv[\rho,\theta,z]=[\sqrt{x^2+y^2},\atantwo(y,x),z]$. The magnitude of $\Mr$ is assumed to be constant: $\Mr=\text{M}_s\mhatr$. A value of M$_s$=\SI{384}{\kA/\meter} is assumed here. 

For the numerical calculations, the cylindrical $\rr$ domain of the Hopfion was discretized with [107,65,71] points in cylindrical coordinates, and the $\pp$ space with 950 points for $|\pp|$ between 0 and 20L$^{-1}$, and with 2592 points for $\phat$ at each $|\pp|$. The multipolar orders from $j=1$ to $j=9$ were considered. 

All the expected outcomes such as the conditions in \Eq{eq:real}, or the vanishing of the momenta and angular momenta, are numerically verified for the Hopfion up to numerical inaccuracies at the level of the fifth significant digit.
\begin{table}
\begin{center}
	{\large
	\begin{tabular}{ c c }
		\hline
		$\langle \text{M}|\Lambda|\text{M}\rangle$ &$\langle \text{M}|\text{J}^2|\text{M}\rangle$\\ 
		-129.1 $\hbar$& 1.30$\times10^{3}$ $\hbar^2$ \\
		\hline
\end{tabular}
	}
\end{center}
	\caption{\label{tab:ljp} Helicity $\Lambda$, and angular momentum squared $\text{J}^2$ stored in a Hopfion computed with the last lines of \Eq{eq:helf} and \Eq{eq:jzjpjm}, respectively. An analytical approximation of the Hopfion in a chiral FeGe magnet of cylindrical shape \cite{Sutcliffe2018} was used in the calculations. The height of the cylinder is equal to the magnetic helical period $\text{L}$, and the diameter is equal to 3$\text{L}$. A magnetization density saturation value of M$_s$=\SI{384}{\kA/\meter} is assumed.}
\end{table}

Table~\ref{tab:ljp} contains the values for the helicity and angular momentum squared stored in the Hopfion computed with the last lines of \Eq{eq:helf} and \Eq{eq:jzjpjm}, respectively. The stored helicity is equivalent to $\approx$ 129 right--handed circularly polarized photons, which is about ten orders of magnitude smaller than the number of photons in a circularly polarized femtosecond laser pulse of \SI{10}{\milli\joule\cm\tothe{-2}} fluence at a central wavelength of \SI{800}{\nano\meter}.

The value of helicity, equal to -129.1$\hbar$, implies a lower bound for the number of circularly polarized photons that would be needed in a helicity--dependent all optical switching of the Hopfion onto its mirror image of opposite helicity: \protect{$\lceil 129.1\times 2 \rceil=259$}. Additionally, \protect{-129.1$\hbar$} also bounds the helicity that can be radiated by the Hopfion as it loses its chirality, for example by the action of a large magnetic bias aligning its magnetization density vector along the same direction at all points.

Appendix~\ref{app:amcontent} contains further analysis of the angular momentum content of the Hopfion. In particular, we show that the Hopfion is an eigenstate of $\text{J}_z$ with angular momentum zero.

\section{Conclusion\label{sec:concl}}

We have introduced a new way to obtain expressions for the computation of the fundamental quantities in static matter from its charge and magnetization densities. The method is based on a scalar product obtained from requirements of invariance under the ten-parameter conformal group in three-dimensional Euclidean space $C_{10}(3)$. This group is obtained as the static ($\omega=0$) restriction of the symmetry group of Maxwell equations with sources, namely, the fifteen-parameter conformal group in 3+1 Minkowski spacetime. 

In an exemplary application, we have used the formalism to compute the angular momentum squared and helicity stored in a Hopfion inside a FeGe disk.

We foresee that this methodology will in particular be useful for the design and analysis of experiments involving the switching between stable states of a material system. 

\begin{acknowledgments}
	This work was funded by the Deutsche Forschungsgemeinschaft (DFG, German Research Foundation) -- Project-ID 258734477 -- SFB 1173, and by the Helmholtz Association via the Helmholtz program ``Materials Systems Engineering'' (MSE).
\end{acknowledgments}

\appendix
\section{The inversion $\RR$ is unitary \label{app:runitary}}

In this appendix we extend some of the results in \cite{Kastrup1970} to prove that the inversion operation $\RR$ is unitary under the scalar product of \Eq{eq:static}[\Eq{eq:fsp}] for $\ff{\lambda}$ meeting:
\begin{equation}
	\langle f|f\rangle=\intdpMsp |\ff{\lambda}|^2 <\infty.
\end{equation}

We first summarize the results in \cite[Section~II]{Kastrup1970}.

While \cite{Kastrup1970} deals mainly with solutions of the scalar wave equation, the results in its Section~II apply in general to functions $\varphi(\pp)$ such that 
\begin{equation}
	\label{eq:finnorm}
	\langle \varphi|\varphi\rangle=\intdpMsp |\varphi(\pp)|^2 <\infty,
\end{equation}
where, instead of the four-momentum $k_\mu$, we just use $\pp$ as the argument since the time component $k_0$ is equal to $|\pp|$ for the $\varphi(k_\mu)$ functions considered in \cite{Kastrup1970}.

Since the inversion $\RR$ commutes with all rotations, the expansion in spherical harmonics $\sphharm{j}{m}{\phat}$
\begin{equation}
	\label{eq:ff}
	\varphi(\pp)=\sum_{j=0}^\infty\sum_{m=-j}^{m=j} \varphi_{jm}(|\pp|)\sphharm{j}{m}{\phat},
\end{equation}
is convenient because the inversion leaves $\sphharm{j}{m}{\phat}$ unchanged. This is a consequence of the preservation of angles by conformal transformations. Only the radial parts $\varphi_{jm}(|\pp|)$ are affected by $\RR$. Radial eigenfunctions
\begin{equation}
	\RR e_{nj}(|\pp|)=(-1)^ne_{nj}(|\pp|)
\end{equation}
are identified in \cite[Equation~11]{Kastrup1970}. They are a complete orthonormal system for functions defined on the positive real line $|\pp|\ge 0$.

The authors of \cite{Kastrup1970} then construct a basis of eigenstates of $\RR$ defined by three integers $(n,j,m)$, which allows one to expand any $\varphi(\pp)$ meeting \Eq{eq:finnorm} as:
\begin{equation}
	\label{eq:enjm}
	\begin{split}
		\varphi(\pp)=\sum_{n=0}^\infty&\sum_{j=0}^\infty\sum_{m=-j}^{m=j} c_{njm}e_{njm}(\pp),\text{ where }\\
		c_{njm}=\langle e_{njm}|\varphi\rangle=&\intdpMsp e_{njm}(\pp)^*\varphi(\pp),\\
		e_{njm}(\pp)=e_{nj}(|\pp|)&\sphharm{j}{m}{\phat},\ \RR e_{njm}(\pp) = (-1)^n e_{njm}(\pp),\\
		\text{ and hence }\RR \varphi(\pp)= &\sum_{n=0}^\infty\sum_{j=0}^\infty\sum_{m=-j}^{m=j} (-1)^n c_{njm}e_{njm}(\pp).
	\end{split}
\end{equation}

The unitary character of $\RR$ under the scalar product
\begin{equation}
	\label{eq:result}
	\langle \varphi|\phi\rangle = \intdpMsp \varphi^*(\pp)\phi(\pp) =\langle \RR\varphi|\RR\phi\rangle
\end{equation}
follows then readily from the third line of \Eq{eq:enjm}, and from $\langle e_{\bar{n}\bar{j}\bar{m}}|e_{njm}\rangle=\delta_{\bar{n}n}\delta_{\bar{j}j}\delta_{\bar{m}m}$.

The result that $\RR$ is unitary can be extended to the $\ff{\lambda}$ functions used in this article as follows.

The angular momentum basis is again convenient for the task. The properties of the Wigner D-matrices imply that the relation inverse to \Eq{eq:ffromF} is
\begin{equation}
	\label{eq:ffl}
		\ff{\lambda}=\sum_{j=1}^\infty\sum_{m=-j}^j \FF{\lambda}\sqrt{\frac{2j+1}{4\pi}}\left[\text{D}_{m\lambda}^j(\phat)\right]^*.
\end{equation}

We see that \Eq{eq:ffl} is very similar to \Eq{eq:ff}. Importantly, the angular and radial dependences are separated into $\sqrt{\frac{2j+1}{4\pi}}\left[\text{D}_{m\lambda}^j(\phat)\right]^*$ and $\FF{\lambda}$, respectively. 

Let us now define the functions:
\begin{equation}
	\label{eq:njml}
e_{njm\lambda}(\pp)=e_{nj}(|\pp|)\sqrt{\frac{2j+1}{4\pi}}\left[\text{D}_{m\lambda}^j(\phat)\right]^*.
\end{equation}

The angular functions $\sqrt{\frac{2j+1}{4\pi}}\left[\text{D}_{m\lambda}^j(\phat)\right]^*$ are, as the $\sphharm{j}{m}{\phat}$, a complete orthonormal system on the sphere (see the solution of the boxed integral in \Eq{eq:yes} of App.~\ref{sec:spam}). Actually, for $\lambda=0$ we have that $\sqrt{\frac{2j+1}{4\pi}}\left[\text{D}_{m0}^j(\phi,\theta,0)\right]^*=\sphharm{j}{m}{\phat}$ \cite[Equation~8.5-10]{Tung1985}. Since the radial functions $e_{nj}(|\pp|)$ are orthonormal and complete on the real positive line, it follows that the $e_{njm\lambda}(\pp)$ in \Eq{eq:njml} are are a basis of eigenstates of $\RR$ with eigenvalue $(-1)^n$, and hence can be readily shown to act unitarily on the spaces of $\ff{\lambda}$ functions by repeating the above considerations around Equations~(\ref{eq:enjm},\ref{eq:result}).

\section{The scalar product in the angular momentum basis\label{sec:spam}}

The properties of the Wigner D-matrices imply that the relation inverse to \Eq{eq:ffromF} is
\begin{equation}
	\label{eq:fffromFF}
	\ff{\lambda}=\sum_{j=1}^\infty\sum_{m=-j}^j \FF{\lambda}\sqrt{\frac{2j+1}{4\pi}}\left[\text{D}_{m\lambda}^j(\phat)\right]^*,
\end{equation}

which we substitute in \Eq{eq:fsp}
\begin{equation}
	\begin{split}
		&\langle \mathrm{M}_1|\mathrm{M}_2\rangle =\sum_{\lambda=-1,0,+1}\intdpMsp \sum_{\bar{j}\bar{m}}\sum_{j m}\sqrt{\frac{2\bar{j}+1}{4\pi}}\sqrt{\frac{2j+1}{4\pi}}\\
		& \left[\prescript{1}{}{\mathrm{f}_{\lambda}^{\bar{j}\bar{m}}(|\pp|)}\right]^*\prescript{2}{}{\FF{\lambda}} \text{D}_{\bar{m}\lambda}^{\bar{j}}(\phat)
\left[\text{D}_{m\lambda}^j(\phat)\right]^*.
	\end{split}
\end{equation}

After splitting the d$^3\pp$ integral into its radial and angular parts 
\begin{equation}
	\label{eq:yes}
	\begin{split}
		&\langle \mathrm{M}_1|\mathrm{M}_2\rangle =\sum_{\lambda=-1,0,+1}\int_{>0}^\infty \ \frac{\text{d}|\pp| |\pp|^2}{|\pp|} \sum_{\bar{j}\bar{m}}\sum_{j m}\sqrt{\frac{2\bar{j}+1}{4\pi}}\sqrt{\frac{2j+1}{4\pi}}\\
		& \left[\prescript{1}{}{\mathrm{f}_{\lambda}^{\bar{j}\bar{m}}(|\pp|)}\right]^*\prescript{2}{}{\FF{\lambda}}\boxed{\int \text{d}^2\phat\ \text{D}_{\bar{m}\lambda}^{\bar{j}}(\phat)\left[\text{D}_{m\lambda}^j(\phat)\right]^*},
	\end{split}
\end{equation}
we solve the angular integral in the box by substituting $\text{D}_{m\lambda}^j(\phat)=\exp(-\ii m\phi)\text{d}^j_{m\lambda}(\theta)$ [\Eq{eq:Dexplicit}], solving the integral in $\phi$, and using the orthogonality properties of the small Wigner d-matrices \cite[Equation~8.3-2]{Tung1985}, whose elements are real-valued:
\begin{equation}
	\label{eq:sp}
	\begin{split}
		&\int_{-\pi}^\pi \text{d}\phi\int_{0}^\pi \text{d}\theta\sin\theta \text{D}_{\bar{m}\lambda}^{\bar{j}}(\phi,\theta,0)\left[\text{D}_{m\lambda}^j(\phi,\theta,0)\right]^*\\
		&=\int_{-\pi}^\pi \text{d}\phi \exp\left(\ii\left(m-\bar{m}\right)\phi\right)\int_{0}^\pi \text{d}\theta\sin\theta\ {\text{d}}_{\bar{m}\lambda}^{\bar{j}}(\theta)\text{d}^j_{m \lambda}(\theta)\\
		&=2\pi \delta_{\bar{m}m} \int_{0}^\pi \text{d}\theta\sin\theta\ {\text{d}}_{\bar{m}\lambda}^{\bar{j}}(\theta)\text{d}^j_{m \lambda}(\theta)\\
		&=\frac{4\pi}{2j+1}\delta_{\bar{m}m}\delta_{\bar{j}j}. 	
	\end{split}
\end{equation}
Substituting this result in the box of \Eq{eq:yes} results in \Eq{eq:Fsp}.

\section{The linear and angular momenta vanish\label{sec:iszero}}

In this appendix we will show that the angular momenta $\langle \mathrm{M}|\text{J}_{\tau \in \{x,y,z\}}|\mathrm{M}\rangle$ and the linear momenta $\langle \mathrm{M}|\text{P}_{\tau\in \{x,y,z\}}|\mathrm{M}\rangle$ vanish by using $\mrest\ \in \mathbb{R}^3$ and its consequences for the $\ff{\lambda}$ and the $\FF{\lambda}$ coefficients.

The well-known consequence of $\mrest\ \in \mathbb{R}^3$ for its Fourier transform is that
\begin{equation}
	\label{eq:mreal}
	\mathbf{M}(\pp)=\left[\mathbf{M}(-\pp)\right]^*,
\end{equation}
which, together with the complex conjugation transformations of the $\mathbf{e}_\lambda(\phat)$ in \Eq{eq:epmez}
\begin{equation}
	\left[\mathbf{e}_{\pm}(\phat)\right]^*=\mathbf{e}_{\pm}(-\phat),\ \left[\mathbf{e}_{0}(\phat)\right]^*=\mathbf{e}_{0}(\phat)=-\mathbf{e}_{0}(-\phat),
\end{equation}
and the decomposition $	\sqrt{\frac{\muz}{\cz\hbar}}\mathbf{M}_\lambda(\pp)= \ff{\lambda}\mathbf{e}_\lambda(\phat)$ in \Eq{eq:mq}, readily lead to

\begin{equation}
	\label{eq:realff}
	\ff{\lambda}=(-1)^{\lambda+1}\left[\text{f}_{\lambda}(-\pp)\right]^*.
\end{equation}

Showing that 

\begin{equation}
	\label{eq:realFF}
	\FF{\lambda}=(-1)^{j-m+1}\left[\text{f}_{j(-m)\lambda}(|\pp|)\right]^*,
\end{equation}

is somewhat more involved. We start by using \Eq{eq:fffromFF} to write:

\begin{equation}
	\label{eq:fstar}
	\left[\text{f}_{\lambda}(-\pp)\right]^*=\sum_{j=1}^\infty\sum_{m=-j}^j\left[\FF{\lambda}\right]^*\sqrt{\frac{2j+1}{4\pi}}\text{D}_{m\lambda}^j(-\phat),
\end{equation}
and manipulate the last term
\begin{equation}
	\label{eq:vars}
	\begin{split}
		&\text{D}_{m\lambda}^j(-\phat)=\exp\left[-\ii m\left(\phi+\pi\right)\right]d_{m\lambda}^j(\pi-\theta)\\
		&=(-1)^m \exp\left(-\ii m \phi\right)(-1)^{j-\lambda}d_{-m\lambda}^j(\theta)\\
		&=(-1)^{j+m-\lambda}\left[ \exp\left(\ii m \phi\right)d_{-m\lambda}^j(\theta)\right]^*\\
		&\duetoeq{eq:Dexplicit}(-1)^{j+m-\lambda}\left[\text{D}_{-m\lambda}^j(\phat)\right]^*,
	\end{split}
\end{equation}
using the change of angular variables upon spatial inversion $\phat\rightarrow -\phat \implies (\theta,\phi)\rightarrow (\pi-\theta, \phi+\pi)$ in the first equality, a formula in \cite[Equation~(1), p.~79]{Varshalovich1988} in the second, and that the small Wigner d-matrices are real valued in the third. 

After substituting \Eq{eq:vars} into \Eq{eq:fstar} 
{\small
\begin{equation}
\left[\text{f}_{\lambda}(-\pp)\right]^*=\sum_{j=1}^\infty\sum_{m=-j}^j(-1)^{j+m-\lambda}\left[\FF{\lambda}\right]^*\sqrt{\frac{2j+1}{4\pi}}\left[\text{D}_{-m\lambda}^j(\phat)\right]^*
\end{equation}
}

and re-labeling the summation variable $m\rightarrow -m$
{\small
\begin{equation}
	\label{eq:varstwo}
	\left[\text{f}_{\lambda}(-\pp)\right]^*=\sum_{j=1}^\infty\sum_{m=-j}^j(-1)^{j-m-\lambda}\left[\text{f}_{j(-m)\lambda}(|\pp|)\right]^*\sqrt{\frac{2j+1}{4\pi}}\left[\text{D}_{m\lambda}^j(\phat)\right]^*,
\end{equation}
}

comparing \Eq{eq:varstwo} with \Eq{eq:fffromFF} makes it clear that the condition $\ff{\lambda}=(-1)^{\lambda+1}\left[\text{f}_{\lambda}(-\pp)\right]^*$ in \Eq{eq:realff} implies \Eq{eq:realFF}.

\subsection{Angular momentum\label{sec:miszero}}

The proof that $\langle \mathrm{M}|\text{J}_z|\mathrm{M}\rangle=0$ can be found in the main text, just below \Eq{eq:real}.

Let us now consider $\langle \mathrm{M}|\text{J}_y-\ii\text{J}_x|\mathrm{M}\rangle$ in \Eq{eq:jzjpjm}, and show that
\begin{equation}
	\sum_{m=-j+1}^{m=j}\sqrt{(j+m)(j-m+1)} \left[\FF{\lambda}\right]^*\mathrm{f}_{j(m-1)\lambda}(|\pp|)=0,
\end{equation}
which implies that $\langle \mathrm{M}|\text{J}_y-\ii\text{J}_x|\mathrm{M}\rangle=0$. 

The summation in $m$ contains an even number of terms since $\mathrm{f}_{j(m-1)\lambda}(|\pp|)$ is not defined for $m=-j$. The summing can be done pairwise, where one of the terms has $m>0$ and the other is the $\bar{m}=-m+1$ term. Dropping elements from the notation, their sum reads:
\begin{equation}
	\begin{split}
		&\sqrt{(j+m)(j-m+1)} {\mathrm{f}_{jm}}^*\mathrm{f}_{j(m-1)}+\\
		&\sqrt{(j+\bar{m})(j-\bar{m}+1)} {\mathrm{f}_{j\bar{m}}}^*\mathrm{f}_{j(\bar{m}-1)}.
	\end{split}
\end{equation}
We now substitute $\bar{m}=-m+1$
\begin{equation}
	\label{eq:kt}
	\begin{split}
		&\sqrt{(j+m)(j-m+1)} {\mathrm{f}_{jm}}^*\mathrm{f}_{j(m-1)}+\\
		&\sqrt{(j-m+1)(j+m)} {\mathrm{f}_{j(-m+1)}}^*\mathrm{f}_{j-m},
	\end{split}
\end{equation}
and find that the sum vanishes after applying \Eq{eq:realFF} to the two $\mathrm{f}$ factors of the second line of \Eq{eq:kt}: 
\begin{equation}
	\begin{split}
		&\sqrt{(j+m)(j-m+1)}[ {\mathrm{f}_{jm}}^*\mathrm{f}_{j(m-1)}+\\
		&{\mathrm{f}_{j(m-1)}}(-1)^{-j-m}{\mathrm{f}_{jm}}^*(-1)^{j+m+1}]=0.
	\end{split}
\end{equation}

The steps for the corresponding proof that \mbox{$\langle \mathrm{M}|\text{J}_y+\ii\text{J}_x|\mathrm{M}\rangle$=0} in the second line of \Eq{eq:jzjpjm} are very similar.

\subsection{Linear momentum\label{sec:piszero}}
	In $\pp$-space, the action of the momentum operator $\text{P}_{\tau\in\{x,y,z\}}$ is $\text{P}_{\tau}\Mp = \hbar k_\tau \Mp$. Then, the momentum of $\Mr$ in direction $\tau$ can be written:
\begin{equation}
	\label{eq:k}
	\begin{split}
		\langle \mathrm{M}|\text{P}_{\tau}|\mathrm{M}\rangle& = \intdpM\left[\Mp\right]^\dagger\hbar k_\tau\Mp\\
		&=\intdpMnohbar k_\tau|\Mp|^2.
	\end{split}
\end{equation}
	Since $\Mr$ is a real-valued field, we have from \Eq{eq:mreal} that $\Mp=\left[\mathbf{M}(-\pp)\right]^*$, hence $|\Mp|^2=|\mathbf{M}(-\pp)|^2$, which readily leads to the conclusion that $\langle \mathrm{M}|\text{P}_{\tau}|\mathrm{M}\rangle=0$ from \Eq{eq:k}. 

\section{Angular momentum content of the Hopfion\label{app:amcontent}}
This appendix contains further analysis of the angular momentum content of the Hopfion. Some insight can be gained even without having the $\FF{\lambda}$ at hand. For example, we now show that the Hopfion magnetization density is an eigenstate of $\text{J}_z$ with eigenvalue zero. 

We start by slightly rewriting the expression in \cite[Equation~(3.3)]{Sutcliffe2018} with the help of a rotation matrix: 
\begin{equation}
	\label{eq:hopfion}
	\begin{split}
		&\mhatr=\begin{bmatrix}\hat{m}_x(\rho,\theta,z)\\\hat{m}_y(\rho,\theta,z)\\\hat{m}_z(\rho,\theta,z)\end{bmatrix}=\\
			&\begin{bmatrix}\cos\theta&-\sin\theta&0\\\sin\theta&\cos\theta&0\\0&0&1\end{bmatrix}\begin{bmatrix}4\Xi\Omega\rho\\4\Xi\left(\Upsilon-1\right)\rho\\(1+\Upsilon)^2-8\Xi^2\rho^2\end{bmatrix}\frac{1}{(1+\Upsilon)^2}\\
				&=R_z(\theta)\mhatthetazero,
	\end{split}
\end{equation}
	where the last equality contains the definition of $\mhatthetazero$, and 
	\begin{equation}
		\begin{split}
			&\Xi=\left(1+\left(2z/\text{L}\right)^2\right)\sec\left(\pi\rho/\left(2\text{L}\right)\right)/\text{L},\\
			&\Omega=\tan\left(\pi z/\text{L}\right), \ \Upsilon=\Xi^2\rho^2+\Omega^2/4.
		\end{split}
	\end{equation}
We now apply apply the $\text{J}_z$ of \cite[Equation~(5.43)]{Rose1957} to the Hopfion. In cylindrical coordinates for $\rr$ but Cartesian components for the vector, $\text{J}_z$ has the following expression:
\begin{equation}
	\label{eq:jz}
	\text{J}_z\equiv -\hbar\ii\partial_\theta + \hbar\begin{bmatrix}0&-\ii&0\\\ii&0&0\\0&0&0\end{bmatrix}.
\end{equation}
Since $\text{M}_s$ does not depend on $\theta$, we can just apply $\text{J}_z$ to $R_z(\theta)\mhatthetazero$:
\begin{equation}
	\label{eq:zero}
	\begin{split}
		&\frac{1}{\hbar}\text{J}_z R_z(\theta)\mhatthetazero=-\ii\partial_\theta R_z(\theta)\mhatthetazero+\begin{bmatrix}0&-\ii&0\\\ii&0&0\\0&0&0\end{bmatrix}
R_z(\theta)\mhatthetazero\\
		&=\begin{bmatrix}\ii\sin\theta&\ii\cos\theta&0\\-\ii\cos\theta&\ii\sin\theta&0\\0&0&1\end{bmatrix}\mhatthetazero+\begin{bmatrix}-\ii\sin\theta&-\ii\cos\theta&0\\\ii\cos\theta&-\ii\sin\theta&0\\0&0&1\end{bmatrix}\mhatthetazero\\
		&=\zerovec.
	\end{split}
\end{equation}

\begin{figure}[t!]
	\includegraphics[width=\linewidth]{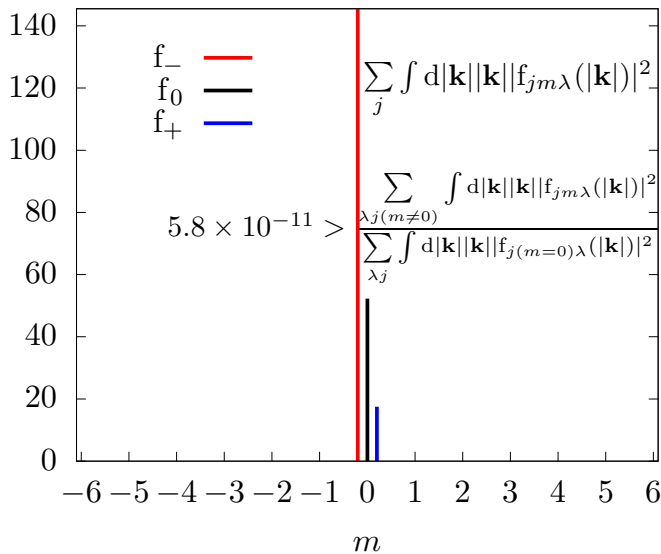}
	\caption{\label{fig:fFHopfion} Angular momentum content of the Hopfion for each helicity computed with the formula on the top right corner. Such expression gives the partial norm squared of the Hopfion in each $(m,\lambda)$ subspace, and is therefore a conformally invariant unitless number. While the numerical errors in the calculations produce $m\neq 0$ components, they are not visible in this scale because they are much smaller than the $m=0$ components, as indicated by the inequality. For clarity, two of the color bars are horizontally offset from the $m=0$ point.}
\end{figure}

The result of \Eq{eq:zero} is faithfully reproduced by the $\FF{\lambda}$ amplitudes obtained from the sequence of computations indicated in \Eq{eq:sequence}. Figure~\ref{fig:fFHopfion} shows the distribution of the different helicities across the values of $m$. We highlight that the values for $m\neq 0$ are not suppressed from the plot, but they are not visible in this scale. The sign of the helicity of the Hopfion in Tab.~\ref{tab:ljp} can be deduced from the larger value of the negative helicity component in Fig.~\ref{fig:fFHopfion}.

\end{document}